\documentstyle[12pt]{article}
\newcommand{\be}{\begin{equation}}
\newcommand{\ba}{\begin{eqnarray}}
\newcommand{\ee}{\end{equation}}
\newcommand{\ea}{\end{eqnarray}}

\begin{document}

\title{The Scattering amplitude for one parameter family of shape invariant 
potentials related to $X_m$ Jacobi polynomials}

\author{ Rajesh Kumar Yadav $^{a}$\footnote{e-mail address: rajeshastrophysics@gmail.com  }, 
Avinash Khare$^{b}$\footnote {e-mail address: khare@iiserpune.ac.in}
 and  {Bhabani Prasad Mandal$^{a}$}\footnote {e-mail address: bhabani.mandal@gmail.com}}
 \maketitle
{$~^a$ Department of Physics,Banaras Hindu University,Varanasi-221005, INDIA.\\ 
$~^b$Raja Ramanna Fellow, Indian Institute of Science Education and Research (IISER),Pune-411021, INDIA.}

\begin{abstract}
We consider the recently discovered, one parameter family of exactly solvable 
shape invariant potentials which are isospectral to the generalized 
P\"oschl-Teller potential. By explicitly considering the asymptotic 
behaviour of the $X_m$ Jacobi polynomials associated with this system 
($m= 1,2,3,...$), the scattering amplitude for the one parameter family of
potentials is calculated explicitly. 

\end{abstract}

\section{Introduction}
 
The ideas of Supersymmetric quantum mechanics (SQM) and
shape invariant potential (SIP) have not only enriched our understanding
of the exactly solvable potentials but have helped in substantially
increasing the list of exactly solvable potentials \cite{cks}. 
In particular, the search for the exactly
solvable potentials has been boosted greatly due to the recent discovery of 
exceptional orthogonal polynomials (EOP) (also known as $X_m$ Laguerre and 
$X_m$ Jacobi polynomials) \cite{eop1,eop2,eop3}. Unlike the usual orthogonal 
polynomials, these EOPs start with degree $m\geq 1$
 and still form a complete orthonormal set with respect to a positive definite 
innerproduct defined over a compact interval. This remarkable 
work lead Quesne \cite{que} to the discovery of two new SIPs (with translation)
 whose solution is in terms of $X_1$ Laguerre and $X_1$ Jacobi polynomials. 
Soon afterwords, a third SIP (with translation) was discovered whose solution 
is also in terms of $X_1$ Jacobi
polynomials \cite{bqr}. Subsequently, Odake and Sasaki constructed three one 
parameter family of shape invariant potentials (with translation) whose bound 
state eigenfunctions
are in terms of $X_m$ Laguerre and $X_m$ Jacobi polynomials \cite{os}. 
It is worth reminding here that all of these are isospectral to the well known
SIPs.
 
It is not usually appreciated that unlike the usual SIPs, the newly discovered 
SIPs are explicitly $\hbar$ dependent. Further, while two out of the three 
newly discovered SIPs have pure bound state spectrum, the third SIP which is 
isospectral to the generalized P\"oschl-Teller (GPT) potential, has both
discrete and continuum spectrum. Recently, we have calculated the scattering 
amplitude for the SIP which is isospectral to GPT potential and whose 
bound state eigenfunction is in terms of $X_1$ Jacobi polynomial \cite{rab}.
The purpose of the present paper is to extend this work to a class of
isospectral potentials. In particular, in this paper we consider one
parameter family of SIPs which are isospectral to GPT and whose bound state
eigenfunction is given in terms of $X_{m}$ Jacobi polynomials ($m=1,2,3...$) 
and obtain the scattering amplitude for this family by considering the 
asymptotic behaviour of the $X_m$ exceptional Jacobi polynomials(EOP).\\ 

This paper is organized as follow: In Sec. $2$, to set the notation, we 
briefly review the work of Odake and Sasaki \cite{os} regarding the bound state 
eigenvalues and eigenfunctions for the one parameter family of SIPs which are 
isospectral to GPT and whose bound state eigenfunctions are in terms of
$X_m$ Jacobi polynomials. To motivate our calculation for the general case, 
the scattering amplitude for the potential with the bound state 
eigenfunction in terms of  $X_2$ Jacobi polynomial
is discussed in Sec. $3$. The most general $X_m$ case is discussed in 
Sec. $4$. We summarize our conclusions in Sec. $5$.

\section{Bound State of Infinitely many shape invariant potentials}

In this section we essentially set the notation by reviewing the work of
Odake and Sasaki \cite{os} regarding one parameter family SIPs
  and the corresponding bound states.
We mostly adopt their notations in this paper. \\
For $m\geq 1$, the shape invariant prepotential $\omega_l(r;{\bf\lambda})$  
which is isospectral to the GPT is given by
\be\label{sp}
\omega_{m}(r;{\bf \lambda} )= \omega_{0}(r;{\bf\lambda}+ m{\bf\delta}) + \log \frac{\xi_{m}(\cosh 2r;{\bf\lambda}+ {\bf\delta})}{\xi_{m}(\cosh 2r;{\bf\lambda})};  \qquad 0\leq r\leq \infty
\ee
where
${\bf\lambda} = (g,h), \ \ h > g > 0, \ \ {\bf\delta} = (1,-1)$, \ \ 
while $\xi_{m}(\cosh 2r;{\bf\lambda})$ is related with the Jacobi polynomial 
as follows:
\be\label{xi}
\xi_{m}(\cosh 2r;{\bf\lambda})= P_{m}^{(-g-m-\frac{1}{2},-h+m-\frac{3}{2})}(\cosh 2r) \,.
\ee
It may be noted that the prepotential related with GPT corresponds to $m=0$ 
and is given by 
\be\label{gp}
\omega_{0}(r;{\bf\lambda}) = g \log \sinh r - h \log \cosh r\,.
\ee
The general Hamiltonian corresponding to the prepotential 
$\omega_m(r;{\bf\lambda})$ is given by  
\be\label{h}
{\cal H}_{m}({\bf\lambda})={\cal A}_{m}({\bf\lambda})^{\dagger}{\cal A}_{m}({\bf\lambda}) =  p^2 + V_m(r)\,,
\ee
where 
\be
 p = -i\partial_r \qquad V_{m}(r) = \omega_m^{'} (r;{\bf\lambda})^2 + \omega_m^{''} (r;{\bf\lambda})\,,
\ee
\be
{\cal A}_{m}({\bf\lambda})= \partial_r - \omega_{m}^{'}(r;{\bf\lambda}), \qquad {\cal A}_{m}({\bf\lambda})^{\dagger}= -\partial_r -\omega_{m}^{'}(r;{\bf\lambda})\,.
\ee
Here prime on $\omega_{l}(r;{\bf\lambda})$ denotes  derivative with respect 
to r. The prepotential $\omega_m (r;{\bf\lambda})$ satisfies the shape invariance 
condition 
\be
 {\cal A}_m({\bf\lambda}){\cal A}_m({\bf\lambda})^\dagger = {\cal A}_m({\bf\lambda}+\delta ){\cal A}_m({\bf\lambda}+\delta )^\dagger + {\cal E}_1({\bf\lambda}+m\delta)
\ee
This further implies
\be
 \omega_m^{'} (r;{\bf\lambda})^2 - \omega_m^{''} (r;{\bf\lambda}) = \omega_m^{'} (r;{\bf\lambda}+{\bf\delta})^2 + \omega_m^{''} (r;{\bf\lambda}+{\bf\delta})+{\cal E}_1({\bf\lambda}+{m\bf\delta}),
\ee
 in which $\delta$ is a certain shift of the parameter $\bf\lambda$.
The general form of entire set of discrete eigenvalues and  
corresponding eigenfunctions of ${\cal H}_{m}({\bf\lambda})$ are obtained by solving, 
\be\label{se}
{\cal H}_{m}({\bf\lambda})\psi_{m,\nu}(r;{\bf\lambda})= {\cal E}_{m,\nu}({\bf\lambda})\psi_{m,\nu}(r;{\bf\lambda})
\ee
The discrete eigenvalues are,
\ba\label{ee}
{\cal E}_{m,\nu}({\bf\lambda}) = {\cal E}_{\nu}({\bf\lambda}+m{\bf\delta})
 = \sum_{k=0}^{\nu-1} {\cal E}_1({\bf\lambda}+k{\bf\delta}+m{\bf\delta}) 
=4\nu(h-g-2m-\nu)\,
\ea
with $ \nu=0,1,2...,\nu_B-m; \qquad \nu_B=\frac{(h-g)}{2}$. The corresponding eigenfunctions are written as 
\ba\label{wfn}
\psi_{m,\nu}(r;{\bf\lambda}) &=& {\phi_{m}(r;{\bf\lambda})}P_{m,\nu}(\cosh 2r;{\bf\lambda}) \nonumber \\
&=& \frac{e^{\omega_{0}(r;{\bf\lambda}+m{\bf\delta})}}{\xi_{m}(\cos h 2r;{\bf\lambda})}P_{m,\nu}(\cosh 2r;{\bf\lambda})\,,
\ea
with
\be\label{poly}
P_{m,\nu}(\cosh 2r;{\bf\lambda})= a_{m,\nu}(r;{\bf\lambda})P_{\nu}(\cosh 2r;{\bf\lambda}+m{\bf\delta})+b_{m,\nu}(r;{\bf\lambda})P_{\nu-1}(\cosh 2r;{\bf\lambda}+m{\bf\delta})\,.
\ee
Here the coefficients $a_{m,\nu}(r;{\bf\lambda})$ and 
$b_{m,\nu}(r;{\bf\lambda})$ are given by \cite{os}
\ba\label{am}
a_{m,\nu}(r;{\bf\lambda}) = \xi_{m}(\cosh 2r;g+1,h-1)&+&\frac{2\nu(-g-h+m-1)\xi_{m-1}(r;g,h-2)}{(-g-h+2m-2)(g-h+2\nu +2m-1)} \nonumber \\
&-&\frac{\nu(-2h+4m-3)\xi_{m-2}(r;g+1,h-3)}{(2g+2\nu+1)(-g-h+2m-2)}\,,
\ea
\be\label{bm}
b_{m,\nu}(r;{\bf\lambda}) = \frac{(-g-h+m-1
 )(2g+2\nu+2m-1)\xi_{m-1}(r;g,h-2)}{(2g+2\nu+1)(g-h+2\nu+2m-1)}\,.
\ee
It is worth noting that the polynomials $P_{m,\nu}(r;{\bf\lambda})$ are 
orthogonal with respect to the measure $\phi_{m}(r;{\bf\lambda})^2$, i.e.
\ba\label{oc}
&&\int_0^{\infty} dr \phi_{m}(r;{\bf\lambda})^{2} P_{m,\nu}(r;{\bf\lambda})P_{m,q}(r;{\bf\lambda})\nonumber \\
&& = h_{m,\nu}(g,h)\delta_{\nu m}= h_{\nu}(g+m,g-m)\frac{(\nu+g+m+\frac{1}{2})(h-\nu-2m+\frac{1}{2})}
{(\nu+g+\frac{1}{2})(h-\nu-m+\frac{1}{2})}\delta_{\nu q}\,, 
\ea
where
\be
h_{\nu}({\bf\lambda})= \frac{\Gamma (\nu +g+\frac{1}{2})\Gamma (h-g-\nu +1)}{2\nu!(h-g-2\nu)\Gamma (h-\nu+\frac{1}{2})}\,. 
\ee
It is remarkable that even though the potential related with GPT, i.e. 
$V_{GPT} = \omega_0^{'} (r;{\bf\lambda})^2 - \omega_0^{''} (r;{\bf\lambda})$ 
is very different from the potential 
$V_m = \omega_m^{'} (r;{\bf\lambda})^2 - \omega_m^{''} (r;{\bf\lambda}) $, the 
bound state spectrum(\ref{ee}) of the two for any integral $m$ is still the 
same, however the corresponding eigenfunctions are different.  
 Replacing $2r$ by $r$, and after using (\ref{xi}),(\ref{gp}) in (\ref{wfn}), 
the bound state wave function  related to the $X_{m}$ Jacobi polynomial, is 
given by
\be\label{wf}
\psi_{\nu}^m(r) = N_\nu ^{m} \frac{(\cosh r - 1)^{\frac{1}{2}(\alpha + 1/2)} (\cosh r + 1)^{\frac{1}{2}(\beta + 1/2)}}{P_{m}^{(-\alpha-1,-\beta-1)}(\cosh r)} \hat{P}_{\nu + m} ^{(\alpha,\beta)}(\cosh r)      
\ee
where \qquad $\alpha=g+m-\frac{1}{2} $ , $\beta=-h+m-\frac{1}{2} $,
 $N_\nu^{m} = [2^{(h-g-2m+1)}h_{m,n}(g,h)]^{\frac{1}{2}}$, is the 
normalization constant, $\hat{P}^{(\alpha , \beta,m )} _{\nu+m}(\cosh r) $ is 
$(\nu+m)$ th-degree  $X_m $ Jacobi Polynomial and 
$P_{m}^{(-\alpha-1,-\beta-1)}(\cosh r)$ is usual Jacobi polynomial. 

\section{Calculation of scattering amplitude for $m$=$2$ ($X_{2}$ Jacobi polynomial)}
The relation between the $X_m$ Jacobi polynomial and the usual Jacobi 
polynomial is given by \cite{hos}\\
\ba\label{xm}
P_{m,\nu}(\cosh r)&=&\hat P_{\nu+m}^{(\alpha,\beta)}(\cosh r)=\left (P_m^{(-\alpha -2,\beta )}(\cosh r)+\frac{2\nu(m- \alpha +\beta -1) P_{m-1}^{- \alpha ,\beta}(\cosh r)}
{(2m -\alpha +\beta -2)(2\nu + \alpha +\beta)}\right. \nonumber \\ 
 &-& \left. \frac{\nu (\beta +m-1) P_{m-2}^{(-\alpha ,\beta )}(\cosh r)}{(\alpha +\nu -m+1)(2m-\alpha +\beta - 2)}  
P_\nu^{(\alpha,\beta)}(\cosh r)\right ) \nonumber \\
&+&\frac{(m-\alpha +\beta-1)(\alpha+\nu )}{(\alpha+\nu-m+1)(2\nu+\alpha+\beta)}P_{m-1}^{(-\alpha,\beta)} (\cosh r) P_{n-1}^{(\alpha,\beta )} (\cosh r) 
\ea
where
\be \label{jp1}
P_\nu^{(\alpha,\beta)}(\cosh r)=\frac{\Gamma(\alpha+\nu+1)}{\nu!(\Gamma(\alpha+\beta+l+1)}\sum_{q=1}^\nu \left( \begin{array}{clcr}
\nu \\
p
\end{array} \right)\frac{\Gamma(\alpha+\beta+\nu+p+1)}{\Gamma(\alpha+p+1)}\left(\frac{\cosh r-1}{2}\right)^q  
\ee
Using this relation, we have recently calculated the scattering amplitude for 
the $m = 1$ ($X_1$ Jacobi case) \cite{rab}. We now extend that discussion
to the $m=2$ case. For $X_2$ Jacobi case, we set $m=2$ in the above expression, to get   
\ba \label{x2}
\hat P_{\nu+2}^{(\alpha,\beta)}(\cosh r)& = &\left[\frac{1}{2}\{\alpha(\beta+2)+(\alpha-\beta-2)(\alpha-\beta-1)\}-\frac{(\alpha-\beta-1)(\beta-\alpha+2)}{8}x^2 \right.  \nonumber \\
&+& \left. \left(\frac{(\alpha-\beta-1)(\alpha+\beta+2)}{4}-\frac{\nu(\beta-\alpha+1)(\alpha-\beta-2)}{(\beta-\alpha+2)(\beta+\alpha+2\nu)}\right)x \right. \nonumber \\
&-& \left. \frac{\nu(\beta-\alpha+1)(\alpha+\beta)}{(\beta-\alpha+2)(\beta+\alpha+2\nu)}-\frac{\nu(\beta+1)}{(\alpha+\nu-1)(\beta-\alpha+2)}\right] P_\nu^{(\alpha,\beta)}(\cosh r) \nonumber \\
&-&\frac{(\beta-\alpha+1)(\alpha+\nu)}{2(\beta+\nu-1)(\alpha+\beta+2\nu)} [(\alpha+\beta-2)x+(\alpha+\beta)] P_{\nu-1}^{(\alpha,\beta)}(\cosh r)
\ea
The usual Jacobi polynomial 
$P^{(\alpha , \beta )} _{\nu} (\cosh r)$ can be written in terms of 
Hypergeometric function as :\\
 \be\label{jp}
P^{(\alpha , \beta )} _{\nu} (\cosh r) = \frac{\Gamma(\nu+\alpha+1)}{ \nu! \Gamma (1+\alpha)} F(\nu+\alpha+\beta+1
, -\nu , 1+\alpha ; \frac{1-\cosh r}{2})\,.           
 \ee
To get the scattering state wave functions for this system, two modifications 
of the bound state wavefunctions are required \cite{ks}:
(i) The second solution of the Schr\"odinger equation which diverges asymptotically and hence had been 
discarded earlier, must be retained.
(ii) The discrete level  $\nu $ should be replaced by the wavenumber $k$ such that one gets asymptotic behavior in terms of $e^{\pm ikr}$ as $r\rightarrow \infty $.\\
 Equation(\ref{jp}) can be written by considering the second solution as,  
 \ba\label{jh1}
P^{(\alpha , \beta )} _{\nu} (\cosh r)& = & \frac{\Gamma (\nu+\alpha+1)}{ \nu ! \Gamma (1+\alpha)} \left[C_1 F(\nu+\alpha+\beta+1
,-\nu , 1+\alpha ; \frac{1-\cosh r}{2})\right. \\ \nonumber 
& + & \left. C_2 (\frac{1-\cosh r}{2})^{-(\nu+\alpha+\beta+1)} \ \ 
F(\nu+\beta + 1, -\nu -\alpha , 1- \alpha ; \frac{1-\cosh r}{2})\right]    
\ea
We consider the boundary condition, $r\rightarrow 0$ , i.e.$(\frac{1-\cosh r}{2}) \rightarrow 0, \psi_{\nu}(r)\rightarrow $ finite, the allowed solution is
\be\label{11}
P^{(\alpha , \beta )} _{\nu} (\cosh r) = \frac{\Gamma(\nu+\alpha+1)}{ \nu ! \Gamma (1+\alpha)} C_1 F(\nu+\alpha+\beta+1
,-\nu , 1+\alpha ; \frac{1-\cosh r}{2})     
\ee
where $C_1$ is a constant.\ \
In order to compare our results with the previous results \cite{rab}, we use  
$\alpha = B-A-\frac{1}{2}$ , $\beta = -B-A-\frac{1}{2}$. 
Now replacing $\nu$ by $A+ik$, we get
\be\label{jh3}
P^{(\alpha , \beta )} _{(A+ik)} (\cosh r) = C_1\frac{\Gamma(B+ik+1/2)}{ (A+ik) ! \Gamma (B-A+1/2)}F(-A+ik,-A-ik,B-A+1/2;\frac{1-\cosh r}{2})\,,          
\ee
\ba\label{jh4}
P^{(\alpha ,\beta )}_{(A+ik-1)}(\cosh r)& = & C_1\frac{\Gamma(B+ik-1/2)}{(A+ik-1)!\Gamma (B-A+1/2)}  \nonumber  \\ 
&& \times  F(-A+ik-1,-A-ik+1, B-A+1/2;\frac{1-\cosh r}{2})\,.  
\ea
Using Eqs. (\ref{jh3}) and (\ref{jh4}) in (\ref{x2}) we get  $\hat P^{(\alpha , \beta )} _{(\nu+2)} (\cosh r) = \hat P^{(\alpha , \beta )} _{(A+ik+2)} (\cosh r)$.\\
The scattering state wavefunction thus  is given by\\ 
$\psi_{k}(r)=$
\be\label{ss}
   N_k^2\frac{(\cosh r - 1)^{\frac{1}{2}(B-A)}(\cosh r + 1)^{-\frac{1}{2}(B+A)}P_{A+ik+1} ^{(\alpha,\beta)}(\cosh r) }{\left( 2(B-1)(2B-1)\cosh^2 r+2(2B+1)(2A+1)\cosh r+4A^2+4A+2B-1\right)} 
\ee
Using the properties of hypergeometric function \cite{toi},
\ba\label{hf}
F(\alpha,\beta,\gamma;z) &=& (1-z)^{-\alpha} \frac{\Gamma (\gamma) \Gamma (\beta - \alpha)}{\Gamma (\beta) \Gamma (\gamma - \alpha)}F(\alpha,\gamma - \beta,        
\alpha-\beta+1; \frac{1}{1-z})  \nonumber   \\
&+&(1-z)^-\beta \frac{\Gamma (\gamma) \Gamma (\alpha - \beta)}{\Gamma (\alpha) \Gamma (\gamma - \beta)}F(\beta,\gamma - \alpha,
\beta-\alpha+1; \frac{1}{1-z})\,,      
\ea
and taking the limit $r\rightarrow \infty$, finally we get the asymptotic
form of (\ref{ss}), as
\be\label{asym1}
\lim_{r\to\infty}\psi_{k}(r) = N_k^2 \frac{C_1 P
2^{A+1}4^{ik}}{16} 
\left[\left( \frac{ac}{P}\right)2^{-4ik} e^{ikr} + e^{-ikr}\right]\,,          
\ee
where\\
\[P = \frac{(B+ik-3/2)(2ik-1)(ab)+2(B+ik-1/2)(ed)}{(B+ik-3/2)(2ik-1)} ; \ \ a = \frac{\Gamma (B+ik+1/2)}{(A+ik)!\Gamma (B-A+1/2)} ;\]\\
\[b = \frac{\Gamma(B-A+1/2)\Gamma(-2ik)}{\Gamma(-A-ik)\Gamma (B-ik+1/2)}; \ \ c = \frac{\Gamma (B-A+1/2)\Gamma (2ik)}{\Gamma (-A+ik)\Gamma (B+ik+1/2)};\]\\
\[d = \frac{\Gamma (B+ik-1/2)}{(A+ik-1)!\Gamma (B-A+1/2)} ;\ \ e = \frac{\Gamma (B-A+1/2)\Gamma (-2ik+2)}{\Gamma (-A-ik+1)\Gamma (B-ik+3/2)};\] 
The asymptotic behavior for the radial wavefunction (for l=0) is given by 
\cite{cks}
\be\label{asym2}
\lim_{r\to\infty}\psi_{k}(r) \simeq \frac{1}{2k}[S_{l=0} e^{ikr} 
 -  e^{-ikr} ]    
\ee
From (\ref{asym1}) and (\ref{asym2}) we get
\be\label{sct1}
S_{l=0} = \left (\frac{ac}{P}\right )2^{-4ik}                      
\ee
Using the values of P, a, b, c, d and e, we obtain the scattering
amplitude for the $X_2$ Jacobi case
\ba\label{sct2}
S_{l=0} &=& S_{l=0}^{GPT} \left[\frac{\{ B^2-(ik-1/2)^2 \} - (B-ik+1/2)}{\{ B^2-(ik+1/2)^2\} -(B+ik+1/2)}\right] \nonumber \\
&=& \frac{\Gamma(2ik)\Gamma(-A-ik)\Gamma(B-ik+1/2) 2^{-4ik}}
{\Gamma(-A+ik)\Gamma(-2ik) \Gamma(B+ik+1/2)}\times \nonumber \\
&&\left[\frac{\{ B^2-(ik-1/2)^2 \} - (B-ik+1/2)}{\{ B^2-(ik+1/2)^2\} -(B+ik+1/2)}\right]\
\ea

\section{Calculation of scattering amplitude for $X_{m}$ Jacobi case}
We now proceed to generalize this calculation to the $X_m$ case 
($m=1,2,3,...$). 
Following the calculation done above for the $X_2$ case, using Eqs. (\ref{jp1})
 and (\ref{11}) in Eq. (\ref{xm}) and replacing 
$\nu\rightarrow A+ik$ we get 
$\hat P^{(\alpha , \beta )} _{(\nu+m)} (\cosh r) = \hat P^{(\alpha , \beta )} _{(A+ik+m)} (\cosh r)$.
Now using $\hat P^{(\alpha , \beta )} _{(A+ik+m)} (\cosh r)$  and 
then taking the limit $r\rightarrow 0$, we obtain the asymptotic form 
of the wave function (\ref{wf}) for the $X_m$ case  
\be\label{asym3}
\lim_{r\to\infty}\psi_{k}(r) = N_k^m \frac{C_1 \Gamma(-2B+m-1)\Gamma(-2B+2m-1)4^{-A-2m+ik} P}{\Gamma(-3B+A+2m-1/2)\Gamma(-2B+m-1)}
\left[\left( \frac{ac}{Q}\right)2^{-4ik} e^{ikr} + e^{-ikr}\right]\,,          
\ee
where
\be
Q = (ab)+\frac{m(m-2B-1)(B+ik-1/2)}{(B+ik-im+1/2)(2ik-1)}(ed)\,,
\ee
while a, b, c, d, and e, are same as in the $X_2 $ case.
From (\ref{asym3}) and (\ref{asym2}) we get
\be\label{sctm}
S_{l=0} = \left (\frac{ac}{Q}\right )2^{-4ik}                      
\ee
Using  Q, a, b, c,d and e as given above, we finally have the expression for 
the scattering amplitude in the $X_m$ case 
\ba\label{sctm2}
S_{l=0} &=& S_{l=0}^{GPT}\left[\frac{\{ B^2-(ik-1/2)^2 \} + (B-ik+1/2)(1-m)}{\{ B^2-(ik+1/2)^2\} +(B+ik+1/2)(1-m)}\right] \nonumber \\
&=& \frac{\Gamma(2ik)\Gamma(-A-ik)\Gamma(B-ik+1/2) 2^{-4ik}}
{\Gamma(-A+ik)\Gamma(-2ik) \Gamma(B+ik+1/2)}\times \nonumber \\
&&\left[\frac{\{ B^2-(ik-1/2)^2 \} + (B-ik+1/2)(1-m)}{\{ B^2-(ik+1/2)^2\} +(B+ik+1/2)(1-m)}\right] 
\ea
As expected, in the special case of $m=1$ and $2$ we get back the expressions
for the scattering amplitude as obtained in \cite{rab} and in Sec. III above,
thereby providing a powerful check on the calculations. Remarkably, in the 
limit $m=0$, the scattering amplitude as given by Eq. (\ref{sctm2}) reduces
to $S_{l=0}^{GPT}$, providing a further check on the calculations.
It is amusing to note that
the as one goes from $m=1$ to arbitrary integer value, there is simply a change
by a factor of $(1-m)$ in the second term in both the numerator and 
the denominator.   
\section{Summary}  
In this paper we have calculated the scattering amplitude for one parameter
family of potentials (isospectral to GPT), whose bound state eigenfunctions 
are given in terms of $X_m$ Jacobi polynomials. The bound state eigenvalues
and eigenfunctions for these potentials were known before \cite{os}. Thus,
with the calculation in this paper, one now has a complete knowledge about both
the bound state spectrum  and the scattering amplitude for the one parameter
family of SIPs which are isospectral to GPT.

{\bf Acknowledgment}

One of us (RKY) acknowledges financial support from UGC under the FIP Scheme.


\begin{thebibliography}{99}
\bibitem{cks}  F. Cooper, A. Khare, U. Sukhatme {\it  Phys. Rep.  } \textbf{251} (1995) 267; {\it  "SUSY in Quantum Mechanics"  } World Scientific (2001).            
\bibitem{eop2} D. Gomez-Ullate, N. Kamran and R. Milson, {\it J. Math. Anal.Appl.} \textbf{359} (2009) 352.  
\bibitem{eop3} D. Gomez-Ullate, N. Kamran and R. Milson, {\it J. Phys. A} \textbf{43} (2010) 434016. 
\bibitem{eop1} B. Midya and B. Roy, {\it Phys. Lett. A}(2009). 
\bibitem{que}  C. Quesne, {\it J.Phys.A} \textbf{41} (2008) 392001.
\bibitem{bqr}  B. Bagchi, C. Quesne and R. Roychoudhary, 
{\it Pramana J. Phys.} \textbf{73}(2009) 337, C. Quesne, SIGMA {\bf 5} (2009)
84; A. Khare, Unpublished.  
\bibitem{hos} C-L. Ho, S ODAKE and R Sasaki, {\it SIGMA} \textbf{7} (2011) 107. 
\bibitem{rab} R. K. Yadav, A. Khare and B. P. Mandal, {\it Annals of Physics} \textbf {331} (2013) 313–316. 
\bibitem{os}  S. Odake and R. Sasaki, {\it Phys. Lett. B}, \textbf{684} 
(2010) 173; ibid {\bf 679} (2009) 414. {\it J. Math. Phys}, \textbf{51}, 053513 (2010).
\bibitem{ks}  A. Khare and Uday P Sukhatme {\it J. Phys. A: Math. Gen } \textbf{21} (1988) L501.
\bibitem{toi} I.S. Gradshteyn, I.M.Ryzhik, and Alan Jeffrey,{\it "Table of Integrals, Series and Products"} Academic Press (1991).
\end{thebibliography}
\end{document}